\documentclass[pss,fleqn]{w-art}
\usepackage{times}
\usepackage{w-thm}
\theoremstyle{plain}

\theoremstyle{definition}

\usepackage[]{graphicx}
\chardef\bslash=`\\ 

\begin{document}
\renewcommand{\copyrightyear}{2005}
\DOIsuffix{theDOIsuffix}
\Volume{XX}
\Issue{1}
\Month{01}
\Year{2005}
\pagespan{1}{}
\Receiveddate{27 September 2005}
\Reviseddate{}
\Accepteddate{}
\Dateposted{}



\title[Weak universality, bicritical points and reentrant transitions]{Weak universality, bicritical points and reentrant transitions in the critical behaviour of a mixed spin-1/2 and spin-3/2 Ising model on the union jack (centered square) lattice}

\author{Jozef Stre\v{c}ka\footnote{E-mail: {\sf jozkos@pobox.sk}, Phone: +421\,55\,6222121\,no.231,
           Fax: +421\,55\,6222124}} 
\address[]{Department of Theoretical Physics and Astrophysics, 
Faculty of Science, \\ P. J. \v{S}af\'arik University,  
Park Angelinum 9, 040 01 Ko\v{s}ice, Slovak Republic}
\begin{abstract}
The mixed spin-1/2 and spin-3/2 Ising model on the union jack lattice is solved 
by establishing a mapping correspondence with the eight-vertex model. It is shown that 
the model under investigation becomes exactly soluble as a free-fermion eight-vertex model
when the parameter of uniaxial single-ion anisotropy tends to infinity. Under this restriction, 
the critical points are characterized by critical exponents from the standard Ising universality class. 
In a certain subspace of interaction parameters, which corresponds to a coexistence surface between 
two ordered phases, the model becomes exactly soluble as a symmetric zero-field 
eight-vertex model. This surface is bounded by a line of bicritical points having interaction-dependent critical exponents that satisfy a weak universality hypothesis.
\end{abstract}

\maketitle                  


\section{Introduction}

Investigation of phase transitions and critical phenomena belongs to the most intensively studied 
topics in the equilibrium statistical physics. A considerable progress in the understanding of 
order-disorder phenomena has been achieved by solving planar Ising models which represent valuable
exceptions of exactly soluble lattice-statistical models with a non-trivial critical behaviour \cite{Bax82}. 
Although phase transitions of planar Ising models have already been understood in many respects 
there are still a lot of obscurities connected with a criticality of more complicated spin systems 
exhibiting reentrant transitions, non-universal critical behaviour, tricritical phenomenon, 
etc. It is worthy to mention, however, that several complicated Ising models can exactly be 
treated by transforming them to the solvable \textit{vertex models}. A spin-1/2 Ising model on 
the union jack (centered square) lattice, which represents a first exactly soluble system exhibiting 
reentrant transitions \cite{Vak66}, can be for instance reformulated as a free-fermion eight-vertex model 
\cite{Wu87}. It should be also pointed out that an equivalence with the vertex models have already provided 
a precise confirmation of the reentrant phenomenon in the anisotropic spin-1/2 Ising models on the 
union jack lattice \cite{Chi87}, generalized Kagom\'e lattice \cite{Aza87} and centered honeycomb 
lattice \cite{Die91} as well.

Despite the significant amount of effort, there are only few exactly soluble Ising models consisting 
of mixed spins of different magnitudes, which are usually called also as {\it mixed-spin Ising models}. 
A strong scientific interest focused on the mixed-spin systems arises partly on account of 
much richer critical behaviour they display compared with their single-spin counterparts and 
partly due to the fact that they represent the most simple models of ferrimagnets having 
a wide potential applicability in practice. Using the extended versions of decoration-iteration 
and star-triangle transformations, Fisher \cite{Fis59} and Yamada \cite{Yam69} derived 
exact solutions of the mixed spin-1/2 and spin-$S$ ($S \geq 1$) Ising models on the honeycomb 
\begin{figure}[htb]
\vspace{0.0cm}
\begin{center}
\includegraphics[width=5cm]{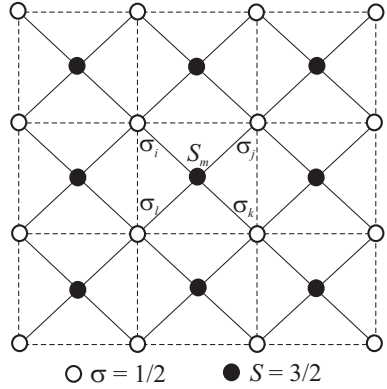}
\end{center}
\vspace{-0.2cm}
\caption{Schematic representation of the the mixed spin-1/2 and spin-3/2 Ising model on the union jack (centered square) lattice. Solid (broken) lines depict the nearest-neighbour (next-nearest-neighbour) interactions.}
\label{fig1}
\end{figure}
and dice lattices, as well as on the decorated honeycomb, rectangular, triangular, and dice lattices. 
Notice that these mapping transformations were later on further generalized in order to account also for the 
single-ion anisotropy effect. The influence of uniaxial single-ion anisotropy have precisely been investigated on the mixed-spin honeycomb lattice \cite{Gon85} and on some decorated planar lattices \cite{Jas98}, while an effect of the biaxial single-ion anisotropy has been explored just on the mixed-spin honeycomb lattice \cite{Str02}. To the best of our knowledge, these are the only mixed-spin planar Ising models with generally known exact solutions except several mixed-spin systems formulated on the Bethe (Cayley tree) lattices, which can be accurately treated within a discrete non-linear map \cite{Sil91} or an approach based on exact recursion equations \cite{Alb03}. 

One of the most outstanding findings to appear in the phase transition theory is being a non-universal critical behaviour of some planar Ising models that is in obvious contradiction with the idea of 
universality hypothesis \cite{Gri70}. The mixed spin-1/2 and spin-1 Ising model on the union jack lattice \cite{Lip95} represents a very interesting system from this viewpoint as it exhibits a remarkable line of bicritical points which have continuously varying critical exponents obeying the \textit{weak universality} hypothesis \cite{Suz74}. Following the approach developed by Lipowski and Horiguchi \cite{Lip95}, i.e. establishing a mapping correspondence with the eight-vertex model, we shall investigate in the present 
article the mixed spin-1/2 and spin-3/2 Ising model on the union jack lattice. In certain subspaces of interaction parameters, the model under investigation becomes exactly soluble either as a free-fermion 
model or a symmetric zero-field eight-vertex model. In the rest of the parameter space one still obtains rather reliable estimate of the criticality within so-called free-fermion approximation \cite{Fan70} 
when a non-validity of the free-fermion condition is simply ignored. 

The outline of this paper is as follows. In Section \ref{sec:model}, a detailed formulation of the model 
system is presented and subsequently, the mapping correspondence that ensures an equivalence with the eight-vertex model will be derived. The most interesting numerical results for a critical behaviour will be presented and particularly discussed in Section \ref{sec:results}. Finally, some concluding remarks are drawn in Section \ref{sec:conclusion}.

\section{Model system and its solution}
\label{sec:model}

Let us begin by considering the mixed spin-1/2 and spin-3/2 Ising model on the union jack (centered square)
lattice ${\cal L}$ as schematically illustrated in Fig. \ref{fig1}. The mixed-spin union jack lattice 
consists of two interpenetrating sub-lattices ${\cal A}$ and ${\cal B}$ that are formed by 
the spin-1/2 (empty circles) and spin-3/2 (filled circles) atoms, respectively. 
The total Hamiltonian defined upon the underlying lattice ${\cal L}$ reads:
\begin{eqnarray}
{\mathcal H}_{mix} = - J  \sum_{(i,j) \subset \mathcal J}^{4N} S_{i} \sigma_{j}
     - J' \sum_{(k,l) \subset \mathcal K}^{2N} \sigma_{k} \sigma_{l}
     - D \sum_{i=1}^{N} S_{i}^2,     
\label{HD}
\end{eqnarray}
where $\sigma_j = \pm 1/2$ and $S_i = \pm 1/2, \pm 3/2$ are Ising spin variables placed on 
the eight- and four-coordinated sites, $J$ denotes the exchange interaction between nearest-neighbouring 
${\cal A}-{\cal B}$ spin pairs and $J'$ labels the interaction between the ${\cal A}-{\cal A}$ spin 
pairs that are next-nearest-neighbours on the union jack lattice ${\cal L}$. Finally, the parameter 
$D$ measures a strength of the uniaxial single-ion anisotropy acting on the spin-3/2 sites and $N$ 
denotes the total number of the spin-1/2 sites. 

\begin{figure}[htb]
\vspace{-0.7cm}
\begin{center}
\includegraphics[width=\textwidth]{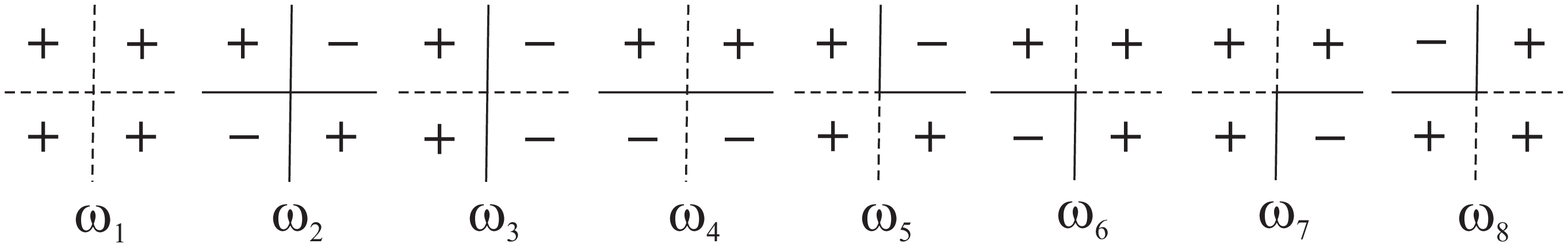}
\end{center}
\vspace{-1.5cm}
\caption{Eight possible spin configurations around each central spin-3/2 site 
and their corresponding line coverings at the vertices of dual square lattice.}
\label{fig2}
\end{figure}

In order to obtain the exact solution, the central spin-3/2 atoms should be firstly decimated from the faces 
of sub-lattice ${\cal A}$. After the decimation, i.e. after performing a partial trace over spin degrees 
of freedom of the spin-3/2 sites (filled circles), the partition function of the mixed-spin union jack lattice ${\cal L}$ can be rewritten as:  
\begin{eqnarray}
{\mathcal Z}_{mix} = \sum_{\{\sigma \}} \prod_{i ,j, k, l} \omega (\sigma_i, \sigma_j, \sigma_k, \sigma_l),    \label{ZD}
\end{eqnarray}
where the summation is performed over all possible spin configurations available on the sub-lattice ${\cal A}$ and the product is over all $N$ faces of the sub-lattice ${\cal A}$, which are constituted by plaquettes composed of a central spin-3/2 site surrounded by four spin-1/2 variables $\sigma_i$, $\sigma_j$, $\sigma_k$, $\sigma_l$ as arranged in Fig. \ref{fig1}. The Boltzmann factor $\omega (a, b, c, d)$ assigned to those faces can be defined as:
\begin{eqnarray}
\omega (a, b, c, d) \! \! \! &=& \! \! \! 2 \exp[\beta J'(ab + bc + cd + da)/2 + \beta D/4]  
\nonumber \\
\! \! \! && \! \! \!
\bigl \{ \exp(2 \beta D) \cosh[3 \beta J (a + b + c + d)/2] + \cosh[\beta J (a + b + c +d)/2] \bigr \},
\label{BF} 
\end{eqnarray}
where $\beta = 1/(k_{\mathrm B} T)$, $k_{\mathrm B}$ is Boltzmann's constant and $T$ stands for 
the absolute temperature. At this stage, the model under investigation can be rather straightforwardly 
mapped onto the eight-vertex model on a dual square lattice ${\cal L_D}$, since the Boltzmann 
factor $\omega (a, b, c, d)$ is being invariant under the reversal of all four spin variables. 
Actually, there are at the utmost eight distinct spin arrangements having different energies 
(Boltzmann weights) and these can readily be related to the Boltzmann weights of the eight-vertex 
model on the dual square lattice. If, and only if, the adjacent spins are aligned opposite to each 
other, then solid lines are drawn on the edges of the dual lattice ${\cal L_D}$, otherwise they are 
drawn as broken lines. Diagrammatic representation of eight possible spin arrangements and their 
corresponding line coverings is shown in Fig.~\ref{fig2}.  
It can be easily understood that there are eight possible line coverings around each vertex of the dual lattice each of them corresponding to two spin configurations, one is being obtained from the other by 
reversing all side spins. Since there is even number of solid (broken) lines incident to each vertex of 
the dual lattice ${\cal L_D}$, the model under consideration becomes equivalent with the eight-vertex model. 
With regard to this equivalence, the partition function of the mixed-spin Ising model on the union jack lattice can be expressed in terms of the partition function of the eight-vertex model on the square lattice:
\begin{eqnarray}
{\mathcal Z}_{mix} (T, J, J', D) = 2 {\mathcal Z}_{8-v} (\omega_1, \omega_2, ..., \omega_8).
\label{PF}
\end{eqnarray}   
The factor 2 in previous equation comes from the two-to-one mapping between spin and 
vertex configurations (two different spin configurations correspond to one vertex configuration). 

The Boltzmann weights, which correspond to eight possible line coverings of the eight-vertex model 
 shown in Fig. \ref{fig2}, can readily be obtained with the aid of equation (\ref{BF}): 
\begin{eqnarray}
\omega_1 \! \! \! &=& \! \! \! 2 \exp(\beta D/4 + \beta J'/2)
                               [\exp(2 \beta D) \cosh(3 \beta J) + \cosh(\beta J)], \nonumber \\
\omega_2 \! \! \! &=& \! \! \! 2 \exp(\beta D/4 - \beta J'/2)[\exp(2 \beta D) + 1], \nonumber \\
\omega_3 \! \! \! &=& \! \! \! \omega_4 = 2 \exp(\beta D/4)[\exp(2 \beta D) + 1], \nonumber \\
\omega_5 \! \! \! &=& \! \! \! \omega_6 = \omega_7 = \omega_8 = 2 \exp(\beta D/4)
                               [\exp(2 \beta D) \cosh(3 \beta J/2) + \cosh(\beta J/2)].  
\label{BW}
\end{eqnarray}   
Unfortunately, there does not exist general exact solution for the eight-vertex model with arbitrary 
Boltzmann weights. However, if the weights (\ref{BW}) satisfy so-called \textit{free-fermion condition}:
\begin{eqnarray}
\omega_1 \omega_2 + \omega_3 \omega_4 = \omega_5 \omega_6 + \omega_7 \omega_8,
\label{FFC}
\end{eqnarray}  
the eight-vertex model becomes exactly soluble as a \textit{free-fermion model} treated several 
years ago by Fan and Wu \cite{Fan70}. It can be readily proved that the free-fermion condition 
(\ref{FFC}) holds in our case just as $D \to \pm \infty$, or $T \to \infty$. The restriction 
to infinitely strong single-ion anisotropy consequently leads to the familiar phase transitions 
from the standard Ising universality class because of the effective reduction of the model system
to a simple spin-1/2 Ising model on the union jack lattice solved many years ago \cite{Vak66, Wu87, Chi87}. Within the manifold given by the constraint (\ref{FFC}), the free-fermion model becomes critical as long as:
\begin{eqnarray}
\omega_1 + \omega_2 + \omega_3  + \omega_4 = 2 \mbox{max} \{ \omega_1, \omega_2, \omega_3, \omega_4 \}.
\label{TCFFC}
\end{eqnarray}
It is noteworthy, however, that the critical condition (\ref{TCFFC}) yields rather reliable 
estimate of the criticality within so-called \textit{free-fermion approximation} \cite{Fan70} 
even if a non-validity of the free-fermion condition (\ref{FFC}) is simply ignored.  

The second branch of exact solution occurs just as the Boltzmann weights (\ref{BW}) satisfy the condition 
of the so-called symmetric zero-field eight-vertex (Baxter) model \cite{Bax82}: 
\begin{eqnarray}
\omega_1 = \omega_2, \quad \omega_3 = \omega_4, \quad \omega_5 = \omega_6, \quad \omega_7 = \omega_8.
\label{S8V}
\end{eqnarray} 
Since we already have $\omega_3 = \omega_4$, $\omega_5 = \omega_6$, and $\omega_7 = \omega_8$, hence,
the symmetric case is obtained by imposing the condition $\omega_1 = \omega_2$ only, or equivalently:
\begin{eqnarray}
\exp(2 \beta D) = \frac{\exp(- \beta J')- \cosh(\beta J)}{\cosh(3 \beta J)-\exp(- \beta J')},
\label{S8V1}
\end{eqnarray}
According to Baxter's exact solution \cite{Bax82}, the symmetric eight-vertex model becomes 
critical on the manifold (\ref{S8V}) if:
\begin{eqnarray}
\omega_1 + \omega_3 + \omega_5  + \omega_7 = 2 \mbox{max} \{ \omega_1, \omega_3, \omega_5, \omega_7 \}.
\label{TCS8V}
\end{eqnarray} 
It is easy to check that $\omega_1$ always represents in our case the greatest Boltzmann weight, thus, 
the condition determining the criticality can also be written in this equivalent form:
\begin{eqnarray}
\exp(\beta_c J'/2) \! \! \! \! \! \! &[& \! \! \! \! \! \! 
\exp(2 \beta _c D) \cosh(3 \beta _c J) + \cosh(\beta _c J)] = \nonumber \\
1 \! \! \! &+& \! \! \! \exp(2 \beta_c D) + 2 \exp(2 \beta_c D) \cosh(3 \beta _c J/2) + 2 \cosh(\beta_c J/2),
\label{TCS8V1}
\end{eqnarray}
where $\beta_c = 1/(k_{\mathrm B} T_c)$ and $T_c$ denotes the critical temperature. It should be stressed, nevertheless, that the critical exponents (with exception of $\delta$ and $\eta$) describing a phase transition of the symmetric eight-vertex model depend on the function $\mu = 2 \arctan(\omega_5 \omega_7/ \omega_1 \omega_3)^{1/2}$, in fact:
\begin{eqnarray}
\alpha = \alpha' = 2 - \frac{\pi}{\mu}, \quad \beta = \frac{\pi}{16 \mu}, \quad \nu = \nu' = \frac{\pi}{2 \mu},
\quad \gamma = \frac{7 \pi}{8 \mu}, \quad \delta = 15, \quad \eta = \frac{1}{4},
\label{CE}
\end{eqnarray} 
For illustrative purposes, let us explicitly evaluate the critical exponent $\beta$ that determines disappearance of the spontaneous order when the critical temperature is approached from below:
\begin{eqnarray}
\beta^{-1} = \frac{32}{\pi} \arctan \biggl\{ \frac{\exp(2 \beta_c D) \cosh(3 \beta_c J/2) + \cosh(\beta_c J/2)}
{[\exp(2 \beta_c D) + 1]^{3/4} [\exp(2 \beta_c D) \cosh(3 \beta_c J) + \cosh(\beta_c J)]^{1/4}} \biggr \}. 
\label{CB}
\end{eqnarray} 

\section{Results and discussion}
\label{sec:results}

\begin{figure}[htb]
\vspace{-0.4cm}
\begin{center}
\includegraphics[width=7cm]{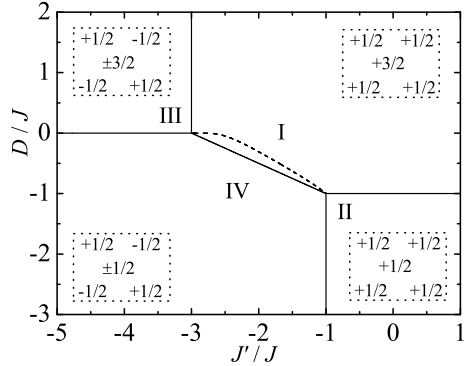}
\end{center}
\vspace{-0.8cm}
\caption{Ground-state phase diagram in the $J'-D$ plane when $J > 0$. Dotted rectangles schematically 
illustrate a typical spin configuration within the basic unit cell of each phase. Broken line connecting both triple points shows a projection of the critical line (\ref{TCS8V1}) into the $J'-D$ plane.}
\label{fig3}
\end{figure}

At first, let us turn our attention to a discussion of the most interesting results obtained for the ground-state and finite-temperature phase diagrams. Solid lines displayed in Fig. \ref{fig3} 
represent ground-state phase boundaries separating four different long-range ordered phases that 
emerge in the ground state when $J > 0$. 
Spin order drawn in dotted rectangles shows a typical spin configuration within the basic unit cell 
of each phase. As could be expected, a sufficiently strong antiferromagnetic next-nearest-neighbour interaction $J'$ alters the structure of the ground state due to a competing effect with the nearest-neighbour interaction $J$. Owing to a competition between the interactions, the central spins are free to flip within the phases III and IV and thus, these phases exhibit a remarkable coexistence of the spin order 
(sub-lattice ${\cal A}$) and disorder (sub-lattice ${\cal B}$). Last but not at least, it is worthwhile 
to mention that a broken line connecting both triple points depicts a projection of the exact critical line (\ref{TCS8V1}) into the $J'-D$ plane. As this projection crosses $T=0$ plane along the ground-state 
transition line $D/J = -3/2 - J'/2J$ between the phases I and IV, it is quite reasonable to suspect 
that this line determines a location of phase transitions between these phases. 

\begin{figure}[htb]
\begin{minipage}[t]{0.48\textwidth}
\includegraphics[width=1.05\textwidth]{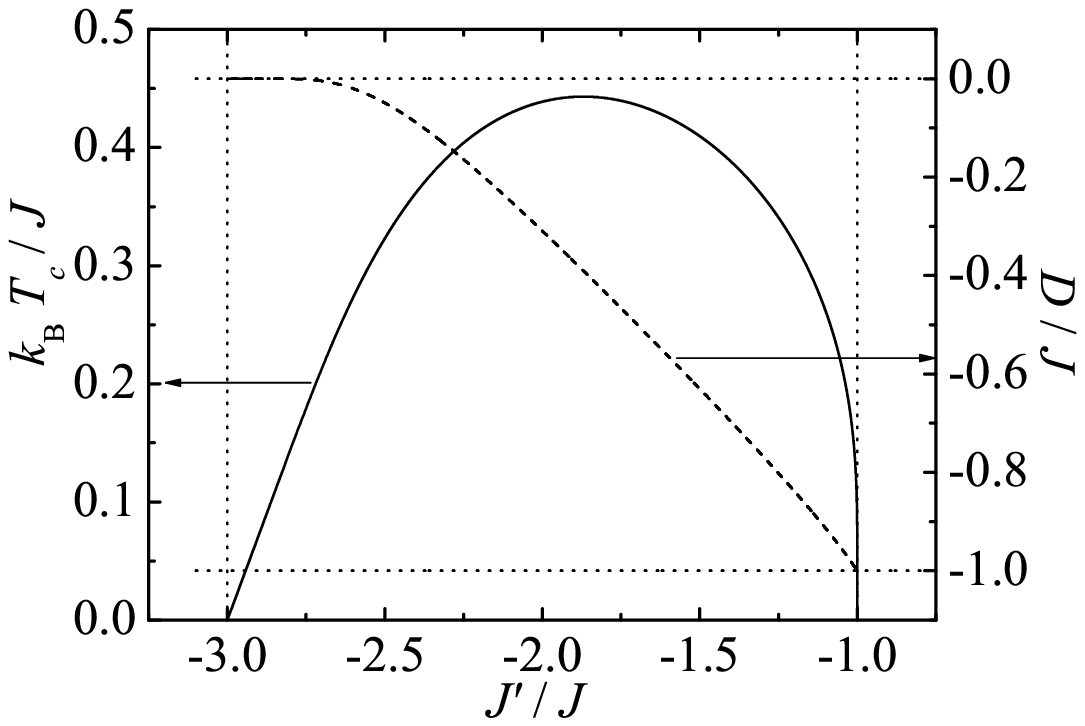}
\vspace{-0.9cm}
\caption{Dependence scaled to a left axis shows how the critical temperature changes with 
the ratio $J'/J$, the curve scaled with respect to a right axis depicts variation of $D/J$ along this line.
Dotted lines are guides for eyes.}
\label{fig4}
\end{minipage}
\hfil
\begin{minipage}[t]{0.48\textwidth}
\includegraphics[width=1.05\textwidth]{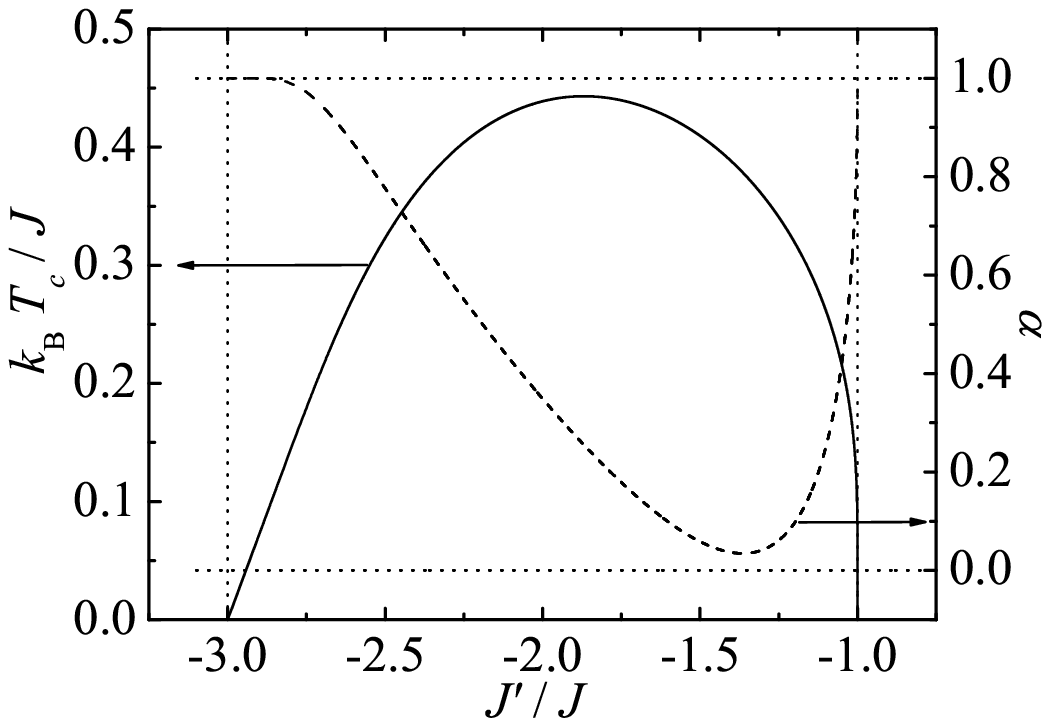}
\vspace{-0.9cm}
\caption{Dependence scaled to a left axis shows how the critical temperature changes with the ratio $J'/J$, the curve scaled with respect to a right axis depicts variations of the critical exponent $\alpha$ along this line.}
\label{fig5}
\end{minipage}
\end{figure}

\begin{figure}[htb]
\begin{minipage}[t]{0.48\textwidth}
\includegraphics[width=1.05\textwidth]{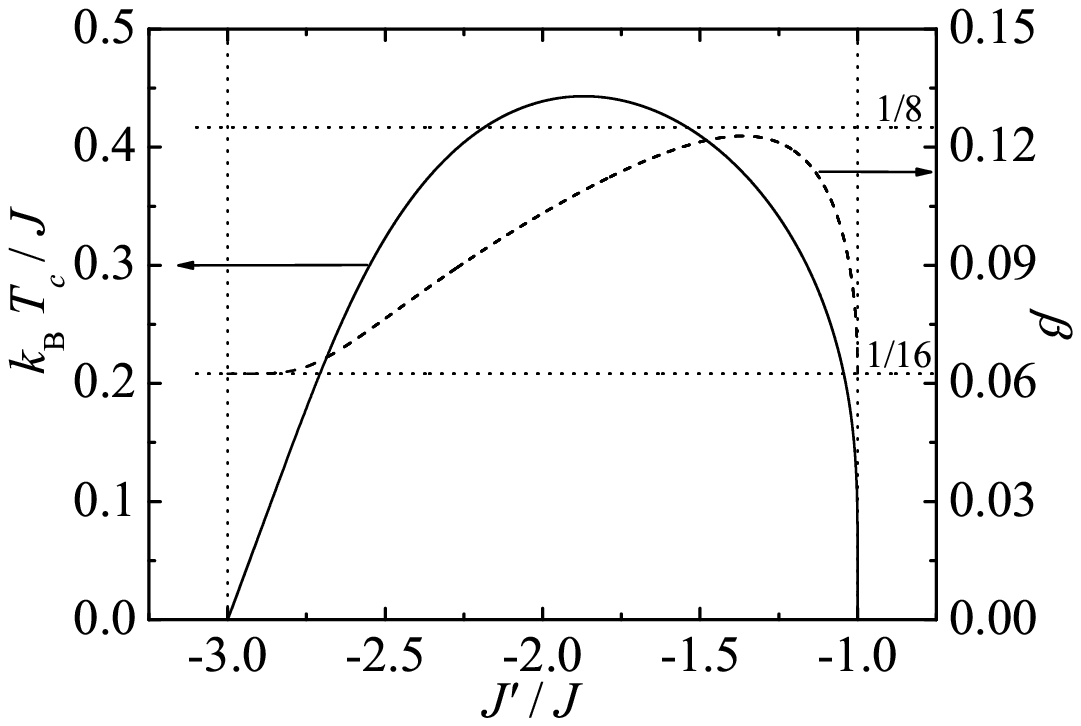}
\vspace{-0.9cm}
\caption{The same as for Fig. \ref{fig5}, but the critical 
index $\beta$ is now scaled with respect to a right axis.}
\label{fig6}
\end{minipage}
\hfil
\begin{minipage}[t]{0.48\textwidth}
\includegraphics[width=1.05\textwidth]{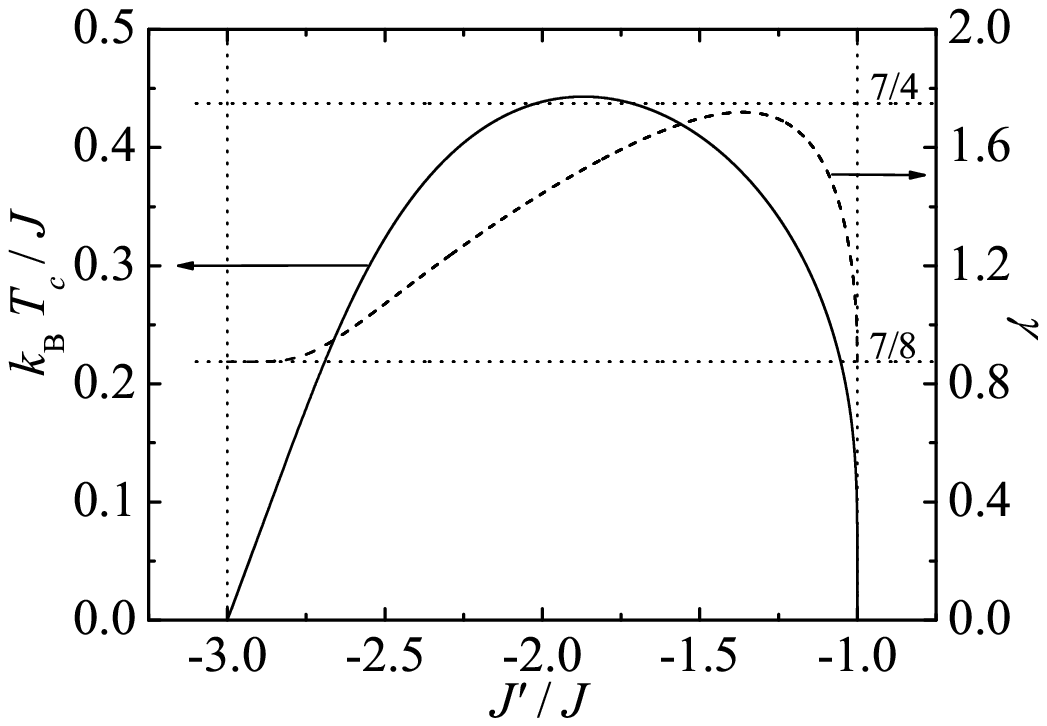}
\vspace{-0.9cm}
\caption{The same as for Fig. \ref{fig5}, but the critical 
index $\gamma$ is now scaled with respect to a right axis.}
\label{fig7}
\end{minipage}
\end{figure}

Let us investigate more deeply this line of critical points. The critical temperatures calculated 
from the symmetric zero-field eight-vertex model must simultaneously obey both the zero-field condition 
(\ref{S8V1}) as well as the critical condition (\ref{TCS8V1}). It is easy to check that the former 
condition necessitates $-3 < J'/J < -1$ and $-1 < D/J < 0$. Fig. \ref{fig4} displays 
a projection of this critical line into the $J'-T_c$ plane (the dependence scaled to the left axis)
and respectively, a projection into the $J'-D$ plane which is scaled to the right axis. Along this 
critical line, the critical exponents are expected to vary with the interaction parameters as they 
have to follow the equations (\ref{CE}). For illustration, Figs. \ref{fig5}, \ref{fig6} and \ref{fig7} 
show how the critical indices $\alpha$, $\beta$, and $\gamma$, respectively, change along the 
critical line. Apparently, the exponents $\beta$ and $\gamma$ approach its smallest  
possible value $1/16$ and $7/8$ by reaching both triple points with zero critical temperature, 
while the critical exponent $\alpha$ approaches there its greatest possible value $1$. 
It is also quite interesting to ascertain that the greatest values for the critical exponents 
$\beta$ and $\gamma$ are slightly below the values $1/8$ and $7/4$, which predicts the 
universality hypothesis for planar Ising systems, while the smallest possible value of 
the critical index $\alpha$ is slightly above its universal value $\alpha \approx 0$ 
(logarithmic singularity). 

\begin{figure}[htb]
\vspace{-0.5cm}
\begin{center}
\includegraphics[width=8cm]{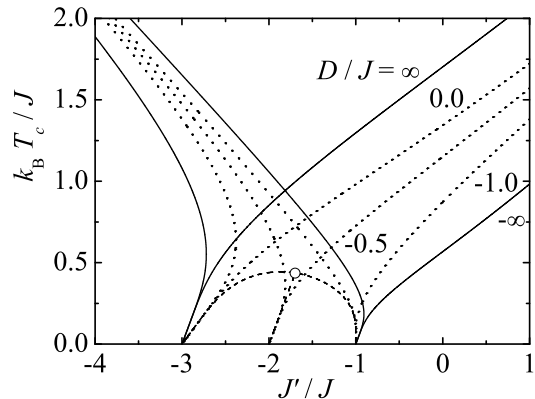}
\end{center}
\vspace{-0.9cm}
\caption{Critical temperature plotted against the ratio $J'/J$ for several values 
of $D/J$. For details see the text.}
\label{fig8}
\end{figure}

Before concluding, few remarks should be addressed to a global finite-temperature phase diagram plotted in Fig.~\ref{fig8}, which displays the critical temperature as a function of the ratio $J'/J$ for several values of the single-ion anisotropy $D/J$. Critical boundaries depicted as solid lines represent exact critical points obtained from the free-fermion solution (\ref{TCFFC}) under the constraint (\ref{FFC})
fulfilled in the limiting cases $D/J \to \pm \infty$. Second branch of exact solution, which is related 
to the critical points of the symmetric eight-vertex model (\ref{TCS8V1}) on the variety (\ref{S8V1}),
is displayed as a rounded broken line. Dotted critical lines show estimated critical temperatures 
calculated from the free-fermion approximation simply ignoring a non-validity of the free-fermion 
condition (\ref{FFC}) for any finite value of $D/J$.

It is quite obvious from the ground-state phase diagram (Fig. \ref{fig3}) that a right (left) 
wing of the displayed critical boundaries corresponds to the phase I (III) if $D/J > 0$, whereas it 
corresponds to the phase II (IV) if $D/J < -1$. Actually, the exact as well as approximate critical 
points resulting from the free-fermion solution correctly reproduce the ground-state 
boundaries between those phases. When the single-ion anisotropy term is selected within the 
range $-1 < D/J < 0$ (see for instance the curve for $D/J = -0.5$), however, both wings are 
expected to meet at a bicritical (circled) point with non-universal (continuously 
varying) critical indices as already reasoned by Lipowski and Horiguchi \cite{Lip95}. In such 
a case, the right and left wings of critical lines separate the phases I and IV, respectively,
and a line of first-order phase transitions is expected to terminate at this special multicritical 
point. There are strong indications supporting this concept \cite{Lip95}, actually, the almost 
straight broken line depicting the zero-field condition (\ref{S8V1}) should show a coexistence 
of these two phases as it starts from a point that determines their coexistence in the ground state. 
In addition, it is quite illuminating to see that both wings of critical temperatures referred to 
the free-fermion approximation start from this ground-state value. With regard to the aforementioned 
arguments one may conclude that a coexistence surface between the phases I and IV lies inside the 
area bounded by the line of bicritical points (rounded broken line) having the non-universal 
critical exponents. 

Finally, we should remark a feasible appearance of the reentrant transitions that can be observed 
in the critical boundaries nearby the coexistence points $J'/J = -3$ and $-1$. It is quite apparent that 
the observed reentrance can be explained in terms of the coexistence of a partial order (sub-lattice
${\cal A}$) and partial disorder (sub-lattice ${\cal B}$) that emerges in both the high-temperature 
reentrant phases III and IV. As a matter of fact, the partial disorder on the sub-lattice ${\cal B}$ 
can compensate a loss of entropy that occurs in these phases as a result of a thermally induced 
partial ordering on the sub-lattice ${\cal A}$ in agreement with a necessary condition 
conjectured for the appearance of reentrant phase transitions \cite{Aza87, Die91}.

\section{Concluding Remarks}
\label{sec:conclusion}

The work reported in the present article provides a relatively precise information on the 
critical behaviour of the mixed spin-1/2 and spin-3/2 Ising model on the union jack lattice 
by establishing a mapping correspondence with the eight-vertex model. The main focus of the present 
work has been aimed at examination of the criticality depending basically on the single-ion anisotropy 
as well as the competing next-nearest-neighbour interaction. The location of the critical boundaries 
has accurately been determined from corresponding solutions of the free-fermion model and the symmetric zero-field eight-vertex model, respectively, whereas the mapping correspondence is being restricted 
to the certain subspaces of interaction parameters where it holds precisely. In the rest of parameter space, 
the free-fermion approximation has been used to estimate the critical boundaries as this method should 
provide rather meaningful approximation giving reliable estimate to the true transition temperatures.

The greatest theoretical interest in the model under investigation arises due to the remarkable 
critical line consisting of bicritical points, which bounds a coexistence surface between two long-range 
ordered phases. The bicritical points can be characterized by the non-universal interaction-dependent 
critical exponents that satisfy the weak universality hypothesis. Moreover, the same arguments 
as those suggested by Lipowski and Horiguchi \cite{Lip95} have enabled us to identify the 
zero-field condition (\ref{S8V1}) with a location of the first-order transition lines separating 
the two ordered phases. 

It should be remarked that the considered system also shows reentrant phase transitions on account of the competition between the nearest- and next-nearest-neighbour interactions. Our results are in agreement 
with the conjecture \cite{Aza87} stating that the reentrance appears as a consequence of the coexistence 
of a partial order and disorder, namely, the partial disorder that appears on the sub-lattice ${\cal B}$ can compensate the loss of entropy which occurs on behalf of the partial ordering on the sub-lattice ${\cal A}$ 
in both the high-temperature partially ordered (disordered) phases.

\begin{acknowledgement}
This work was financially supported under the grants VEGA 1/2009/05 and APVT 20-005204.
\end{acknowledgement}


\begin{thebibliography}{50}

\bibitem{Bax82} 
R.~J.~Baxter, Ann. Phys. \textbf{70}, 193 (1972); \\   
R.~J.~Baxter, Proc. R. Soc. Lond.~A \textbf{404}, 1 (1986);\\   
R.~J.~Baxter, Exactly solved models in statistical mechanics (Academic Press, New York, 1982). 

\bibitem{Vak66}
V.~Vaks, A.~Larkin, and Yu.~N.~Ovchinnikov, Sov. Phys. JETP \textbf{22}, 820 (1966).

\bibitem{Wu87}
F.~Y.~Wu and K.~Y.~Lin, J. Phys.~A: Math. Gen. \textbf{20}, 5737 (1987).

\bibitem{Chi87}
T.~Chikyu and M.~Suzuki, Progr. Theor. Phys. \textbf{78}, 1242 (1987).

\bibitem{Aza87}
P.~Azaria, H.~T.~Diep, and H.~Giacomini, Phys. Rev. Lett. \textbf{59}, 1629 (1987); \\  
M.~Debauche, H.~T.~Diep, P.~Azaria, and H.~Giacomini, Phys. Rev.~B \textbf{44}, 2369 (1991). 

\bibitem{Die91}
H.~T.~Diep, M.~Debauche, and H.~Giacomini, Phys. Rev.~B \textbf{43}, 8759 (1991); \\ 
H.~T.~Diep, M.~Debauche, and H.~Giacomini, J. Magn. Magn. Mater. \textbf{104-107}, 184 (1992).

\bibitem{Fis59}
M.~E.~Fisher, Phys. Rev. \textbf{113},  969 (1959); \\
I.~Syozi, Phase Transitions and Critical Phenomena, Vol. 1, \\
edited by C.~Domb and M.~S.~Green (New York, Academic, 1972).

\bibitem{Yam69}
K.~Yamada, Progr. Theor. Phys. \textbf{42}, 1106 (1969).

\bibitem{Gon85}
L.~L.~Gon\c{c}alves, Phys. Scripta \textbf{32}, 248 (1985); \textbf{33}, 192 (1986); \\
J.~W.~Tucker, J. Magn. Magn. Mater. \textbf{95}, 133 (1999); \\
A.~Dakhama and N.~Benayad, J. Magn. Magn. Mater. \textbf{231}, 117 (2000).

\bibitem{Jas98}
M.~Ja\v{s}\v{c}ur, Physica~A \textbf{252}, 217 (1998); \\
A.~Dakhama, Physica~A \textbf{252}, 225 (1998); \\
S.~Lackov\'a and M.~Ja\v{s}\v{c}ur, Acta Phys. Slovaca \textbf{48}, 623 (1998).

\bibitem{Str02}
J.~Stre\v{c}ka and M.~Ja\v{s}\v{c}ur, Acta Electrotechnica et Informatica \textbf{2}, 102 (2002); \\
J.~Stre\v{c}ka and M.~Ja\v{s}\v{c}ur, Phys. Rev.~B \textbf{70}, 014404 (2004); \\
M.~Ja\v{s}\v{c}ur and J.~Stre\v{c}ka, Physica~A \textbf{358}, 393 (2005).

\bibitem{Sil91}
N.~R.~da Silva, S.~R.~Salinas, Phys. Rev.~B \textbf{44}, 852 (1991).

\bibitem{Alb03}
E.~Albayrak and M.~Keskin, J. Magn. Magn. Mater. \textbf{261}, 196 (2003); \\
E.~Albayrak, Int. J. Mod. Phys.~B \textbf{18}, 3959 (2004); \\
E.~Albayrak and A.~Alci, Physica~A \textbf{345}, 48 (2005); \\
C.~Ekiz, Physica~A \textbf{347}, 353 (2005); \textbf{353}, 286 (2005). 

\bibitem{Gri70}
R.~B.~Griffiths, Phys. Rev. Lett. \textbf{24}, 1479 (1970). 

\bibitem{Lip95}
A.~Lipowski and T.~Horiguchi, J. Phys.~A: Math. Gen. \textbf{28}, L261 (1995); \\
A.~Lipowski, Physica~A \textbf{248}, 207 (1998).

\bibitem{Suz74}
M.~Suzuki, Progr. Theor. Phys. \textbf{51}, 1992 (1974). 

\bibitem{Fan70}
C.~Fan and F.~Y.~Wu, Phys. Rev. \textbf{179}, 560 (1969); \\
C.~Fan and F.~Y.~Wu, Phys. Rev.~B \textbf{2}, 723 (1970). 

\end{thebibliography}
\end{document}